\documentclass{epl}
\usepackage{amssymb,amsmath}

\title{Maximum flow and topological structure of complex networks}
\shorttitle{Maximum flow in complex networks}
\author{Deok-Sun Lee \and  Heiko Rieger}
\shortauthor{Deok-Sun Lee \etal}

\institute{Theoretische Physik, Universit\"{a}t des Saarlandes, 66041 Saarbr\"{u}cken, Germany}


\begin{document}

\maketitle

\begin{abstract}
The problem of sending the maximum amount of flow $q$ between two
arbitrary nodes $s$ and $t$ of complex networks along links with unit
capacity is studied, which is equivalent to determining the number of
link-disjoint paths between $s$ and $t$. The average of $q$ over all
node pairs with smaller degree $k_{\rm min}$ is $\langle
q\rangle_{k_{\rm min}} \simeq c\, k_{\rm min}$ for large $k_{\rm min}$
with $c$ a constant implying that the statistics of $q$ is related to
the degree distribution of the network.  The disjoint paths between
hub nodes are found to be distributed among the links belonging to 
the same edge-biconnected component, and $q$ can be estimated by the number of pairs of
edge-biconnected links incident to the start and terminal node. The
relative size of the giant edge-biconnected component of a network 
approximates to the coefficient $c$. The applicability of 
our results to real world networks is
tested for the Internet at the autonomous system level.
\end{abstract}

The analysis of network structures of complex systems has turned out
to be extremely useful in exploring their large-scale organization and
unveiling their evolutionary origin
~\cite{watts98,albert99,huberman99,jeong00}. Anomalous features found
by their graph-theoretical analyses~\cite{albert02,dorogovtsev02,newman03} 
are the fingerprints of their hidden organization principles as well as 
the key to predicting their behaviors. Real-world networks usually serve a
particular purpose which frequently can be expressed in terms of flow
problems, in particular in the context of transport between the nodes
along links (edges) with restricted capacities. For instance if one is
interested in the maximum possible flow that can be sent from one node
to another,  one has to solve a maximum flow problem, for which
polynomial algorithms exist. It also arises {\it per se} in
a variety of situations such as assignment and scheduling problems
\cite{ahuja93}. In real-world networks one is often interested in the 
question of how many link-disjoint paths do exist between a particular
node pair - for instance to establish as many independent
transportation routes as possible between two nodes in the occasion of
a sudden demand (like in the event of a natural catastrophe or in a
military context). This problem is again identical to the maximum flow
problem between those two nodes in the same network with unit-capacity
links.

Properties of complex networks related to transport are inherently
connected to their topological structure, in particular to various
aspects of connectedness. The existence of a single path from one node
to another is guaranteed if both nodes belong to the same connected
component of the network. Moreover, nodes that belong to a single
biconnected component have two disjoint paths between them. We will
show in this paper that the arrangement of the biconnected components
of a given network is the essential determinant for the number of
disjoint paths between two nodes of large degree, the hubs (which is
non-trivial as soon as more than two disjoint paths between nodes exist).
In this way we also establish a specific connection between the
topological structure and the flow properties of complex networks. 

To be specific we consider undirected networks of
heterogeneous connectivity pattern and calculate the number of
link-disjoint paths between two nodes $s$ and $t$. We  
assign a unit flow capacity to each link and compute numerically
the maximum flow, denoted by $q_{s,t}$, between $s$ and $t$
\cite{hartmann01}. Since only one unit of flow can be sent along 
each link, $q_{s,t}$ is a non-negative integer and is equal to the
number of link-disjoint paths connecting $s$ and $t$.

We will show that on average the maximum flow between two nodes is
proportional to the minimum of their degree (= number of incident
links), with a proportionality constant $c$ that is asymptotically independent 
of this degree. As a consequence we obtain that in scale-free (SF) networks,  
a class of networks where the degree $k$ follows a power-law distribution 
$P_d(k)\sim k^{-\gamma}$ asymptotically, 
the maximum flow obeys a power-law distribution 
with the exponent $2\gamma-1$ while in completely random networks it
follows a Poisson distribution.  Moreover we will demonstrate that the
coefficient $c$ reflects the effect of the global connectivity pattern on 
transport properties such that the constant $c$ varies depending on the total 
number of links as well as the degree exponent.  Finally we will show that
the edge-biconnectedness is crucial in the transport among hub nodes
and that the maximum flow between hub nodes can be estimated by the
number of pairs of edge-biconnected links incident to the start and
terminal node. This implies also that the coefficient $c$ is related to
the relative size of the giant edge-biconnected component.

The degree of a node $i$ is the number of incident links and is denoted as 
$k_i$. For illustrational purposes we first consider $D>1$-dimensional
regular lattices where all nodes have the same degree $2D$ except for
those at the boundary.  Due to the homogeneous connectivity pattern, 
the maximum flow or the number of link-disjoint 
paths between two distinct nodes $s$ and $t$ is $2D$ unless either
node is at the boundary. If $s$ or $t$ is at the boundary, the maximum 
flow is given by the smaller value of $k_s$  and $k_t$. 
Therefore $q_{s,t}$  in regular lattices can be expressed as 
\begin{equation}
q_{s,t} = k_{\rm min}, 
\label{eq:qregular}
\end{equation}
where $k_{\rm min} = \min\{k_s,k_t\}$.  

\begin{figure}
\onefigure[width=0.6\columnwidth]{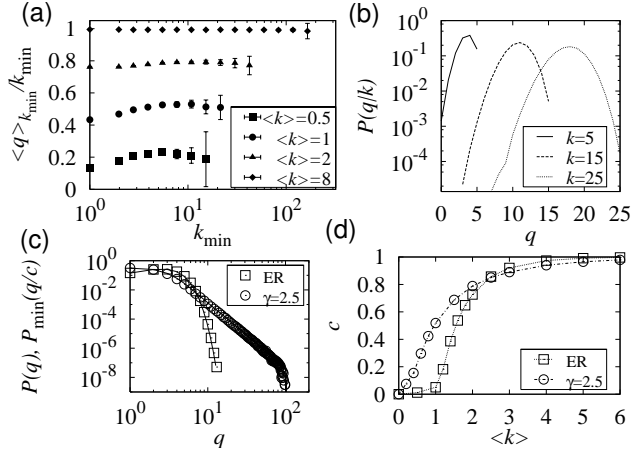}
\caption{Maximum flow in model networks. 
    (a) $\langle q\rangle_{k_{\rm min}}/k_{\rm min}$ 
    versus $k_{\rm min}$ in SF networks 
    with $\gamma=2.5$ and different values of $\langle k\rangle=2L/N$. 
    (b) Conditional distributions $P(q|k)$ 
    in SF networks with $N=1000$, $\langle k\rangle=1.6$, and $\gamma=2.5$.
    (c) Distribution of the maximum flow $P(q)$ for the
    ER graphs and SF networks with $\gamma=2.5$, both 
    consisting of $1000$ nodes and $2000$ links ($\langle k\rangle=4$). 
    Each line represents $P_{\rm min}(q/c)$ with $c\simeq 0.97$ used 
    for the ER graphs and $c\simeq 0.94$ for the SF networks, 
    which is in good agreement with $P(q)$. 
    (d) Flow efficiency $c$ versus $\langle k\rangle$ for the 
    ER graphs and SF networks with $\gamma=2.5$. The statistical error 
    is smaller than the data point size and $N=1000$ in both networks.}
\label{fig:qkmin}
\end{figure}

In contrast, real-world networks typically have a heterogeneous
connectivity pattern: Hub nodes with a very large degree as well as
isolated nodes are simultaneously present. The following question then
naturally arises: Does eq.~(\ref{eq:qregular}) hold for 
such heterogeneous networks? 
To answer this question, we computed the maximum flow for each 
pair of nodes in random SF networks~\cite{lee04} using the 
${\rm MAX\_FLOW}$ algorithm~\cite{leda}. 
Hundreds of network realizations 
were generated for given numbers of nodes $N$ and links $L$, 
and the degree exponent $\gamma$, for which we were able to identify that 
the average of $q$ over the node pairs that have $k_{\rm min}$ as smaller 
degree, $\langle q\rangle_{k_{\rm min}}$, satisfies a relation 
that is analogous to eq.~(\ref{eq:qregular}):
\begin{equation}
\langle q\rangle_{k_{\rm min}} \simeq c \, k_{\rm min},
\label{eq:qrandom}
\end{equation}
for large $k_{\rm min}$, with a coefficient $c$ less than $1$ 
[See fig.~\ref{fig:qkmin}(a)].

If we define the distribution of the maximum flow
as $P(q)\equiv \langle \sum_{s\ne t} \delta_{q_{s,t},q}/ 
(\sum_{s\ne t} 1)\rangle$ with 
$\langle \cdots \rangle$ denoting the ensemble average, 
it can be decomposed in terms of the smaller degree 
as $P(q)= \sum_k P_{\rm min}(k)P(q|k)$, 
where $P_{\rm min}(k)\equiv\langle \sum_{s\ne t}
\delta_{\min\{k_s,k_t\},k}/ (\sum_{s\ne t} 1)\rangle$ and 
$P(q|k) \equiv\langle \sum_{s\ne t}\delta_{q_{s,t},q} \;
\delta_{\min\{k_s,k_t\},k}\rangle/ \langle\sum_{s\ne t}
\delta_{\min\{k_s,k_t\},k}\rangle$.  The conditional distribution
$P(q|k)$ is sharply peaked around the average $q=\langle
q\rangle_{k}$ as shown in fig.~\ref{fig:qkmin}(b), which implies 
$P(q)\approx P_{\rm min}(k=q/c)$ with a good accuracy, where
$c$ is the coefficient appearing in eq.~(\ref{eq:qrandom}).
$P_{\rm min}(k)$ is related to the degree distribution $P_d(k)$ by 
$ P_{\rm min}(k) = 2 P_d(k) \sum_{k'\geq k} P_d(k')$.  Thus
the asymptotic behavior of the degree distribution determines
the large-$q$ behavior of $P(q)$ [fig.~\ref{fig:qkmin}(c)]: 
\begin{equation}
P(q) \sim e^{-({\rm const.})\; q \ln (q/\langle k\rangle)},
\label{eq:pqer}
\end{equation}
for the Erd\H{o}s-R\'{e}nyi (ER) graph 
that have $P_d(k) \sim e^{-k \ln (k/\langle k\rangle)}$~\cite{er61},
and
\begin{equation}
P(q) \sim q^{-(2\gamma-1)},
\label{eq:pqsf}
\end{equation}
for SF networks with the degree exponent $\gamma$. 
Recently L\'{o}pez {\it et al.} reported 
the same asymptotic behavior for the conductance distribution in 
complex networks~\cite{lopez04}.

\begin{figure}
\onefigure[width=\columnwidth]{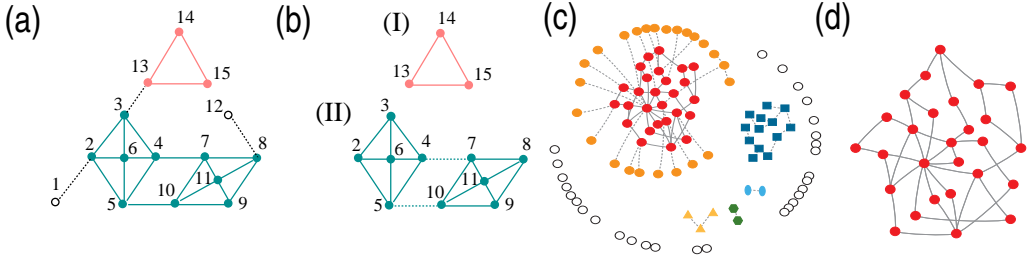}
\caption{(color online)
  {\it Bridge}, {\it separation pair}, and EBCC. 
  (a) The links $(1,2)$, $(8,12)$ and $(3,13)$ are {\it bridges} that 
  would increase the number of CCs if removed~\cite{gabow00}. The network 
  in (a) has two EBCCs (I) and (II) shown in (b). 
  The pair of links $\{(4,7), (5,10)\}$ is a {\it separation pair} 
  that disconnects the EBCC (II). 
  An example of the EBCC in a SF network is shown in (c), which has 
  $100$ nodes, $80$ links, 
  and $\gamma=2.5$.  It has $5$ CCs that has more than one node and 
  the largest CC has an EBCC presented in (d).}
\label{fig:vis}
\end{figure}

Contrary to $D>1$-dimensional regular lattices, even when
$k_{\rm min}$ is large, the maximum flow in heterogeneous networks 
may be very small due to the
collision of connecting paths at some critical links. For example, in
fig.\ref{fig:vis}(a), the maximum flow between the node $2$ and $14$
is only $1$ because every connecting path has to pass the link
$(3,13)$. The value of $c$ in eq.~(\ref{eq:qrandom}) 
  is thus a measure for how efficiently 
such collisions are avoided in the underlying path structure.
From now on, we call the coefficient $c$ {\it flow
efficiency}, which depends on the network topology as shown in
fig.~\ref{fig:qkmin}(d). While $c$ is higher with more links, 
its dependence on the degree exponent $\gamma$ is non-trivial. 
The example in fig.~\ref{fig:qkmin}(d) shows that 
$c$ is higher in SF networks  for $\langle k\rangle\lesssim 2.5$ 
while it is higher in ER graphs otherwise. 
This illuminates ambivalent effects of heterogeneity on transport. 
When links are abundant, 
most node pairs can be connected and have as many 
connecting paths as $k_{\rm min}$. 
What matters is then to avoid collisions of those connecting paths, which can 
be achieved more efficiently in networks closer to a regular one.
On the contrary, when links are deficient but 
as many pathways as possible are required between each pair 
of connected nodes, 
concentration of links within hub nodes is   preferable
to forming a chain of nodes all of which have degree $2$.  

In the following we will show that the flow efficiency is related to
particular topological properties of the network that can be described
by various aspects of connectedness. One unit of flow can be sent from
a node $s$ to another node $t$ if and only if a link $e_{sv}$ incident to $s$ and
a link $e_{tw}$ incident to $t$ belong to the same connected component
(CC) or equivalently, both links are connected. A CC of a graph is a
maximal subgraph in which at least one connecting path exists between each
pair of nodes. Next, the second unit of flow requires another path
that does not share any link with the first path.  This condition is
fulfilled if and only if at least two links $e_{sv}$ and $e_{sv'}$ that are
incident to $s$ and at least two links $e_{tw}$ and $e_{tw'}$ incident
to $t$ belong to the same edge-biconnected component (EBCC) or
equivalently, there exist two pairs of edge-biconnected links,
$(e_{sv},e_{tw})$ and $(e_{sv'},e_{tw'})$. An EBCC is a maximal subgraph which
cannot be disconnected by removing a single link. Each node of a graph
either belongs to a unique EBCC or does not belong to any EBCC.

One could now naively expect that $k>2$ units of flow between $s$ and
$t$ require that $s$ and $t$ belong to
the same $k$-edge-connected component (a $k$-edge-connected graph is
a maximal subgraph which cannot be disconnected by removing any $(k-1)$
links~\cite{gross99} and thus, a graph is $(k-1)$-edge-connected if it
is $k$-edge-connected). This is a sufficient condition, but not a necessary 
one for $k>2$ units of flow:
The nodes on $k$ disjoint paths connecting $s$ and $t$ may have degree $2$ 
and thus the nodes $s$ and $t$ may belong to different $k$-edge-connected
components even though they have $k>2$ link-disjoint paths.

It turns out that already the structure of the connected and the
biconnected components determines the existence of flow values
$q_{s,t}>1$ between two nodes $s$ and $t$. Obviously for $q_{s,t}\ge1$, 
$s$ and $t$ must belong to the same CC. For $q_{s,t}>1$,  $s$ and $t$
should have two pairs of edge-biconnected links. When $s$ and $t$
belong to the same CC but to different EBCCs, the maximum flow will
only be one, $q_{s,t}=1$. There exist one or more {\it
bridges}~\cite{gabow00} between different EBCCs belonging to the same
CC, which would disconnect the EBCCs if removed
[fig.~\ref{fig:vis}]. Analogously if there exists a {\it separation
pair } of links that disconnect $s$ and $t$ if removed
[fig.~\ref{fig:vis}], $q_{s,t}$ will not be larger than $2$ even when
$s$ and $t$ belong to the same EBCC and $k_{\rm min}>2$.  

To study this relation between flow and topology
statistically, we determined all CCs and EBCCs 
in a given network and 
defined for all node pairs $(s,t)$
\begin{equation}
\Theta_{s,t} \equiv \sum_{n=1}^{N_{\rm c}} \min\{f^{(n)}_s,f^{(n)}_t\},
  \quad 
\Sigma_{s,t} \equiv \sum_{m=1}^{N_{\rm ebc}} \min\{g^{(m)}_s,g^{(m)}_t\}.
\label{eq:thetasigma}
\end{equation}
Here $N_{\rm c}$ ($N_{\rm ebc}$) is the total number of CCs (EBCCs) in
a network, $f^{(n)}_i$ ($g^{(m)}_i$) is the number of links incident
to node $i$ and belonging to the $n$-th CC ($m$-th EBCC).
$\Theta_{s,t}$ ($\Sigma_{s,t}$) counts the number of disjoint pairs of
connected (edge-biconnected) links incident to $s$ and $t$. While
$f^{(n)}_{s}$ ($f^{(n)}_{t}$) is either $0$ or $k_s$ ($k_t$),
$g^{(m)}_{s}$ ($g^{(m)}_{t}$) can take any integer between $0$ and
$k_s$ ($k_t$) since each link incident to a node may belong to an EBCC or not.
The deviation of $q_{s,t}$ from $\Theta_{s,t}$ or 
$\Sigma_{s,t}$ is therefore a measure of the number of 
critical links, {\it bridges} or {\it separation pairs}.

\begin{figure}
\onefigure[width=0.75\columnwidth]{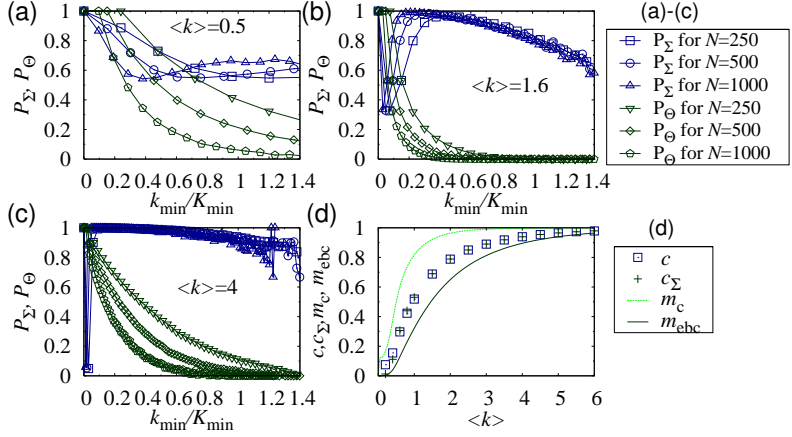}
\caption{(color online) Relation between the maximum flow and edge-biconnectedness. 
  (a)-(c) Conditional probabilities $P_{\Sigma}$ and $P_{\Theta}$ 
  versus $k_{\rm min}/K_{\rm min}$ in 
  SF networks with $\gamma=2.5$ and (a) $\langle k\rangle=0.5$, 
  (b) $\langle k\rangle=1.6$, and (c) $\langle k\rangle=4$. 
  Here $K_{\rm min}$ is the ensemble-averaged largest value of 
  $k_{\rm min}$. Note that $P_{\Sigma}$
  approaches one for increasing average connectivity 
  $\langle k\rangle$. (d) Plots of $c$, $c_{\Sigma}$, $m_{\rm c}$, 
  and $m_{\rm ebc}$ versus $\langle k\rangle$ for SF networks with 
  $\gamma=2.5$ and $N=1000$.} 
\label{fig:cp}
\end{figure}

We computed the conditional probability $P_{\Theta}$ that $q_{s,t}$ is
equal to $\Theta_{s,t}$ and $P_{\Sigma}$ that $q_{s,t}$ is equal to
$\Sigma_{s,t}$ for given values of $k_{\rm min}$ in various model
networks.  It turns out that $\Sigma_{s,t}$ is in a good agreement
with $q_{s,t}$ for large values of $k_{\rm min}$, manifested via
values for $P_{\Sigma}$ close to one.  In SF networks with
$\gamma=2.5$, as shown in fig.~\ref{fig:cp}(a)-(c), $P_{\Sigma}$ is
larger than $0.5$ ($\langle k\rangle=0.5$), $0.8$ ($\langle
k\rangle=1.6$), and $0.9$ ($\langle k\rangle=4$) for a wide range of
$k_{\rm min}/K_{\rm min}\lesssim 1$ independent of the value of
$N$.  $K_{\rm min}$ denotes the ensemble-averaged largest value of
$k_{\rm min}$ that can be observed in networks of $N$ nodes, and is
given by~\cite{lee04} $K_{\rm min}= \langle k\rangle (\gamma-2)
(N/2)^{1/(\gamma-1)}/(\gamma-1)$. Whereas $P_{\Sigma}$ approaches one
for a range of $k_{\rm min}$ that broadens with increasing average
connectivity $\langle k\rangle$, the probability $P_{\Theta}$ is
very small except for the regime where $k_{\rm min}$ is small. 
The agreement of $q_{s,t}$ and $\Sigma_{s,t}$ for large $k_{\rm min}$ 
is observed also for other values of $\gamma>2$ and ER graphs 
($\gamma\to\infty$).

These results elucidate non-trivial features of the paths connecting
hub nodes. Most dangerous links that may prevent large flow between
hub nodes are {\it bridges}. If two hub nodes belong to the same EBCC,
they can send and receive a flow nearly as large as the number of
links belonging to that EBCC. In other words, {\it separation pairs}
are very rare in the EBCC to which hub nodes belong.  The agreement
between the maximum flow and the number of pairs of edge-biconnected
links can be of importance in practical aspects as well: The algorithm
to compute EBCCs has running time $\mathcal{O}(N+L)$~\cite{gabow00}
while the maximum flow takes $\mathcal{O}(N^3)$ time in sequential
machines~\cite{goldberg88}.

The agreement of $q_{s,t}$ and $\Sigma_{s,t}$ allows a deeper understanding
of the relation (\ref{eq:qrandom}). Considering that
$\Sigma_{s,t}$ is dominated by the giant EBCC, defined here as the   
EBCC that has a $\mathcal{O}(L)$ links,   
one can evaluate $\langle \Sigma\rangle_{k_{\rm min}}$, 
the average of $\Sigma$ over the node pairs having $k_{\rm min}$ 
as their smaller degree.
Consider the relative size of the giant EBCC $m_{\rm ebc}$
defined as the ratio (Number of links in the giant EBCC)/(Total
number of links). Assuming that the links of a node
participate in the giant EBCC statistically independently of one another, 
the number of links that belong to the giant EBCC, of $s$ and $t$,
denoted by $g^{(1)}_s$ and $g^{(1)}_t$, respectively, follow Binomial
distributions $B(k_s,m_{\rm ebc})$ and $B(k_t,m_{\rm ebc})$
respectively, giving
\begin{eqnarray}
\langle \Sigma\rangle_{k_{\rm min},k_{\rm max}} 
&\simeq& 
\sum_{g_1=0}^{k_{\rm min}} 
\sum_{g_2=0}^{k_{\rm max}} 
\binom{k_{\rm min}}{g_1}
\binom{k_{\rm max}}{g_2} 
{m_{\rm ebc}}^{g_1+g_2} 
\nonumber\\
&\times&
(1-m_{\rm ebc})^{k_{\rm min} + k_{\rm max} -g_1-g_2} 
\min\{g_1,g_2\}, 
\label{eq:qapprox}
\end{eqnarray}
where $k_{\rm max} = \max\{k_s,k_t\}$ and 
$\langle \cdots\rangle_{x,y}$ means the restricted average 
over the pairs of nodes that have $\min\{k_s,k_t\}=x$ and $\max\{k_s,k_t\}=y$.
Dominant contributions to the summations in eq.~(\ref{eq:qapprox}) 
come from  the regime $(g_1,g_2)\simeq 
(m_{\rm ebc} k_{\rm min}, m_{\rm ebc} k_{\rm max})$ where 
$\min\{g_1,g_2\} = m_{\rm ebc} k_{\rm min}$. 
Thus we have  
\begin{equation}
\langle \Sigma\rangle_{k_{\rm min}} \simeq m_{\rm ebc} \, k_{\rm min}.
\label{eq:q1}
\end{equation}
Equation (\ref{eq:q1}) and the agreement of $q_{s,t}$ and 
$\Sigma_{s,t}$ provide the basis from which eq.~(\ref{eq:qrandom})
follows. In accordance with eq.~(\ref{eq:q1}), 
we indeed observed that $\langle \Sigma\rangle_{k_{\rm min}}
\simeq c_{\Sigma}\, k_{\rm min}$  for large $k_{\rm min}$,  
and the values of $c_{\Sigma}$ and $c$ are compared with 
$m_{\rm ebc}$ in fig.~\ref{fig:cp}(d).
Although $m_{\rm ebc}$ is not exactly equal to $c$ or $c_{\Sigma}$, both 
of which are in excellent agreement, because of 
the correlations among  different links participating in the giant EBCC, 
the deviation is so small that $m_{\rm ebc}$ can be a guide to 
the value of $c$. 
We can also consider $m_{\rm c}$ defined as the 
ratio (Number of links in the giant CC)/(Total
number of links), but it turns out to deviate considerably 
from the value of $c$ or $c_{\Sigma}$ as shown in fig.~\ref{fig:cp}(d).

\begin{figure}
\onefigure[width=\columnwidth]{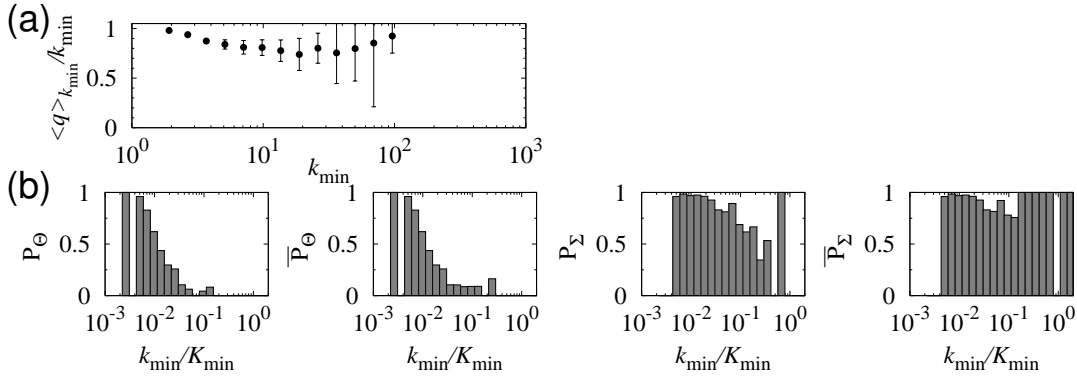}
\caption{Maximum flow for the autonomous system network of the Internet 
  with  $N=5238$, $L=9993$, and $\gamma=2.2(1)$.
  (a) $\langle q\rangle_{k_{\rm min}}/k_{\rm min}$ as a function of 
   $k_{\rm min}$. It fluctuates around $c=0.79(6)$, 
   which is consistent with the relative size of the giant EBCC $m_{\rm ebc}
  \simeq 0.80$.
  (b) Plots of conditional probabilities  $P_{\Theta}$, $\overline{P}_{\Theta}$, $P_{\Sigma}$, and $\overline{P}_{\Sigma}$ versus 
      $k_{\rm min}/K_{\rm min}$ with $K_{\rm min}\simeq 449$.}
\label{fig:real}
\end{figure}

Finally we have tested our findings in a real network, 
the Internet at the autonomous system (AS) level.
An AS includes a set of routers following 
a common routing strategy and the Internet may be viewed 
as a network of the ASs~\cite{yook02}.
Here we used the data recorded on a particular date (June 15, 1999) 
and present the results in fig.~\ref{fig:real}. 
The network consists of $5238$ nodes and $9993$ links 
($\langle k\rangle\simeq 3.82$) 
and the degree exponent is $\gamma=2.2(1)$.
All edge capacities are set to one. 
While all the nodes are connected with one another ($m_{\rm c}=1$), 
only $3251$ nodes and $8006$ links 
belong to the giant EBCC, giving $m_{\rm ebc}\simeq 0.80$.
We found that the relation in eq.~(\ref{eq:qrandom}) is observed and 
furthermore, the flow efficiency $c=0.79 (6)$ is very close to 
the value of $m_{\rm ebc}$. 
The conditional probability $P_{\Sigma}$ is shown to be 
very high in contrast to $P_{\Theta}$. 
Additionally we computed the relaxed probabilities, 
 $\overline{P}_{\Sigma}$ and $\overline{P}_{\Theta}$, 
defined as 
the probability that $|1-q_{s,t}/\Sigma_{s,t}|<0.05$ and 
the probability that $|1-q_{s,t}/\Theta_{s,t}|<0.05$, respectively. 
$\overline{P}_{\Sigma}$ is near one  for most values of $k_{\rm min}$, 
but $\overline{P}_{\Theta}$ differs only slightly from $P_{\Theta}$. 

In conclusion, we studied the relation between transport, specifically 
the maximum flow, and the structural organization in
complex networks. The path structure of complex networks with
heterogeneous connectivity is efficiently organized such that nodes
with more links can send and receive larger amount of flow as expressed
in eq.~(\ref{eq:qrandom}). This may explain the abundance of
heterogeneous connectivity pattern in complex systems that might have
evolved towards better performance of transportation and communication
among their constituents. The structure of
edge-biconnected components determines essentially the maximum flow
between hub nodes, which therefore can be estimated in a time that is
linear in the total number of nodes.

\acknowledgments
We thank H. Jeong for providing the Internet data.

\end{document}